\documentstyle[aps,preprint]{revtex}

\begin{document}
\bibliographystyle{alpha}

\begin{titlepage}

\begin{center}

{\bf {\LARGE Lyapunov Exponents for the Intermittent Transition to Chaos}}

\vspace{.5cm}

{\large
{\bf James Hanssen$^{a}$} and  {\bf Walter Wilcox}$^{a,b}$ \\

\vspace{.5cm}

$^{a}$Department of Physics, Baylor University, Waco, TX 76798 \\

\vspace{.3cm}

$^{b}$Department of Physics, University of Kentucky, Lexington, KY 40506}

\end{center}

\begin{abstract}
The dependence of the Lyapunov exponent on the closeness parameter, $\epsilon$,
in tangent bifurcation systems is investigated. We study and illustrate two
averaging procedures for defining Lyapunov exponents in such systems. First, we
develop theoretical expressions for an isolated tangency channel in which the
Lyapunov exponent is defined on single channel passes. Numerical simulations
were done to compare theory to measurement across a range of
$\epsilon$ values. Next, as an illustration of defining the Lyapunov exponent on many
channel passes, a simulation of the intermittent transition in the logistic map
is described. The modified theory for the channels is explained and a simple
model for the gate entrance rates is constructed. An important correction due to
the discrete nature of the iterative flow is identified and incorporated
in an improved model. Realistic fits to the data were made for the Lyapunov
exponents from the logistic gate and from the full simulation. A number of
additional corrections which could improve the treatment of the gates are identified
and briefly discussed.

\end{abstract}

\vfill
\end{titlepage}
\noindent
{\Large
{\bf 1. Introduction and Background}}
\vspace{.5cm}

Chaos is the study of dynamical systems that have a sensitive dependence on
initial conditions. Much attention has been paid to the two main routes to
chaos: pitchfork bifurcation and tangent bifurcation. If we consider the general
difference equation mapping,
\begin{eqnarray}
x_{n+1}=F(x_{n}),\label{recur}
\end{eqnarray}
then tangent bifurcation, also called type I 
intermitency [Pomeau \& Manneville, 1980], occurs when a tangency develops in
iterates of $F(x_{n})$ across the $x_{n}=x_{n+1}$ reflection line. (Pitchfork
bifurcations occur when iterates of $F(x_{n})$ possess perpendicular crossings of
this line.) Just before the tangency occurs (characterized by the closeness
parameter, $\epsilon$, being small), the map is almost tangent to the reflection
line and a long channel is formed. When the iterations enter, a long laminar-like
flow is established, with nearly periodic behavior. Once the iterations leave the
channel, they behave chaotically, then re-enter the channel. The result is a long
region of laminar flow that is intermittently interrupted by chaotic intervals. This
occurs when $\epsilon$ is near zero and tangency is about to occur, hence the two
names: intermittent chaos and tangent bifurcation. Experimentally, type I
intermittency has been observed in turbulent fluids [Berg\'e {\it et.\ al.}, 1980],
nonlinear oscillators [Jeffries \& Perez, 1982], chemical reactions [Pomeau {\it
et.\ al.}, 1981], and Josephson junctions [Weh \& Kao, 1983]. An excellent
introduction to the intermittency route to chaos is given in Schuster [1995].

In the pioneering studies [Manneville \& Pomeau, 1979] and [Pomeau \&
Manneville, 1980], it was found that the number of iterations followed an
$\epsilon^{-1/2}$ dependence and that the Lyapunov exponent varied as
$\epsilon^{1/2}$ for a logistic mapping ($z=2$). In the work by [Hirsch
{\it et.\ al.}, 1982], an expression for the number of iterations spent
inside the channel was developed. The equation for the third iterate, i.e.
$F(F(F(x)))$ or $F^{(3)}(x)$ where $F(x)=Rx(1-x)$, was expanded in a Taylor series
about one of the tangency points for $R_{c}=1+\sqrt{8}$. In the case of the logistic
map, we get
\begin{eqnarray}
F^{(3)}(x)= x_{c}+(x-x_{c})+a_{c}(x-x_{c})^{2}+b_{c}(R_{c}-R),\label{f3}
\end{eqnarray}
where $x_{c}$ is one of the three contact points. After a transformation
that centers and rescales the system around $x_{c}$ ($y_{n}\equiv
\frac{x_{n}-x_{c}}{b_{c}}$), the recursion relation can be put into the form
\begin{eqnarray}
y_{n+1}=ay_{n}^{2}+y_{n}+\epsilon,\label{what}
\end{eqnarray}
where $\epsilon\equiv R_{c}-R>0$ and $a\equiv a_{c}b_{c}$. The more general case can
be studied as a first, second, or any iterate instead of just the third iterate, as
long as a tangency develops. To derive an analytic description of the trajectory,
[Hirsch {\it et.\ al.}, 1982] switched from a difference equation to a differential
equation. Thus, they considered
\begin{eqnarray}
\frac{dy}{dn}=ay^{2}+\epsilon.\label{diff}
\end{eqnarray}
This approximation is justified as long as the number of iterations
in the channel is large enough or, alternately, that the step size between
iterations is small compared to the channel length. This is an easy differential
equation to solve. One obtains
\begin{eqnarray}
n(y_{in})=\frac{1}{\sqrt{a\epsilon}}\left[
\tan^{-1}\left( y_{out}\sqrt{\frac{a}{\epsilon}}\,\right) -\tan^{-1}\left( y_{in}
\sqrt{\frac{a}{\epsilon}}\,\right) \right].\label{firstn}
\end{eqnarray}
\lq\lq $y_{in}$\rq\rq\, is the entrance to the tangency channel and
\lq\lq $y_{out}$\rq\rq\, is the exit value and one has that
\begin{eqnarray}
-y_{out} \le y_{in} \le y_{out}.\label{limits}
\end{eqnarray}
[Hirsch {\it et.\ al.}, 1982] observed that the entrance points for the
logistic map, $y_{in}$ ($R_{c}\ge R$), had a probability distribution that was
roughly uniform. Given this distribution, the average number of iterations to
travel the length of the channel is given as
\begin{eqnarray}
<n>\equiv
\frac{1}{2y_{out}}\int_{-y_{out}}^{y_{out}}n(y_{in})dy_{in}
=\frac{1}{\sqrt{a\epsilon}}\,\tan^{-1}\left(
y_{out}\sqrt{\frac{a}{\epsilon}}\,\right).\label{it}
\end{eqnarray}
[Hirsch {\it et.\ al.}, 1982] also derived a form for the average number of
iterations for an arbitrary universality class. The universality class, $z$, is
given by the lowest non vanishing power of $(x-x_{c})$ in the expansion around
the tangency point. For tangency to develop, $z$ must always be an even number:
\begin{eqnarray}
y_{n+1}=ay_{n}^{z}+y_{n}+\epsilon.\label{genz}
\end{eqnarray}
This leads to the differential equation,
\begin{eqnarray}
\frac{dy}{dn}=ay^{z}+\epsilon.\label{genzdiff}
\end{eqnarray}
and to the number of iterations,
\begin{eqnarray}
n(y_{in})=a^{-1/z}\epsilon^{-1+1/z}\int\limits_{y_{in}
\sqrt[z]{\frac{a}{\epsilon}}}^{y_{out}\sqrt[z]{\frac{a}{\epsilon}}}\frac{d{\bar
y}}{{\bar y}^{z}+1}.\label{n}
\end{eqnarray}
The average number of iterations is given by
\begin{eqnarray}
<n>=\frac{1}{2}a^{-1/z}\epsilon^{-1+1/z}\int\limits_{-y_{out}
\sqrt[z]{\frac{a}{\epsilon}}}^{y_{out}\sqrt[z]{\frac{a}{\epsilon}}}\frac{d{\bar
y}}{{\bar y}^{z}+1},\label{avgn}
\end{eqnarray}
when the entrance distribution is again uniform. The numerical simulations in
[Hirsch {\it et.\ al.}, 1982] agreed with predicted values quite well.

There are two manners in which Lyanunov exponents may be defined in a
simulation with many trajectories. One may define a procedure which measures the
Lyanunov exponent on a given trajectory, for example a single channel pass, and then
averages over these trajectories. Another possibility is to measure the exponent
across many trajectories or channel passes, using a binning procedure to define
variances. We will use both procedures here to illustrate the theory. The first
procedure will be termed a {\it single pass} measurement, the second a
{\it many pass} measurement. We will develop the theory for the first procedure in
the next Section, which will then be illustrated in Section 3 by a simulation in an
isolated tangency channel for general $z$. As an illustration of a many pass
measurement, a simulation of the intermittent transition in the logistic map will be
described in Section 4. The modified theory will be motivated and a
simple phenomenological model of the data will then be given in Section 5. In
Section 6 an improved expression for the inverse number density, $\frac{dy}{dn}$ due
to the discrete nature of the iterative flow will be developed. This will improve the
comparison of the model to measurement. Finally, we will summarize our findings and
make suggestions for further improvements in the model in the final Section.

\vspace{1cm}
\noindent
{\Large
{\bf 2. Tangency Channel Theory}}
\vspace{.5cm}

Our analysis of the system described by Eqs.(\ref{genz}) and (\ref{genzdiff})
is built on the work of both [Pomeau and Manneville, 1980] and [Hirsch {\it et.\
al.}, 1982]. In contrast to the situation for the average number, $<n>$, little
work has been done to develop expressions for the Lyanupov exponents for
the tangency channel in intermittent systems. We are interested in understanding the
$\epsilon$ dependence of the Lyapunov exponent for $z=2$, finding the
constant of proportionality, and generalizing the results for an arbitrary
universality class, $z$.

The Lyapunov exponent is a measurement which characterizes the sensitive 
dependence on initial conditions of chaotic systems. It is
defined as the coefficient of the average exponential growth per unit time
between initial and final states of a system, which in this case we will take to
be a single pass through a tangency channel. It is given in the case of the
one-dimensional mappings considered here by [Scheck, 1994]
\begin{eqnarray}
\lambda \equiv \lim_{n\to
\infty}\frac{1}{n}\sum_{i=1}^{n}\ln|\frac{dF(y_{i})}{dy_{i}}|.\label{lydef}
\end{eqnarray}
This gives us our starting point for deriving a theory for the Lyapunov 
exponent for a system with an arbitrary universality class. In that case, the
function $F(y_{i})$ from (\ref{genz}) is
\begin{eqnarray}
F(y_{i})=ay_{i}^{z} +y_{i}+\epsilon,\label{feqn}
\end{eqnarray}
so
\begin{eqnarray}
\frac{dF(y_{i})}{dy_{i}}=1 +azy_{i}^{z-1}.\label{fdiff}
\end{eqnarray}
Since we are interested in the Lyapunov exponent for the tangency channel, 
there are only a finite number of steps during which the iterations are confined
to the channel and the appropriate value for n for this trajectory is the total
number of iterations in the channel, $n(y_{in})$. With this in mind, the Lyapunov
exponent is modeled by
\begin{eqnarray}
\lambda(y_{in}) \equiv \frac{1}{n(y_{in})}\int\limits_{y_{in}}^{y_{out}}dn \ln|1
+azy^{z-1}|,\label{lyexp1}
\end{eqnarray}
where we have replaced the discrete sum by an integral over $n$-space. Again,
this step is justified as long as the number of iterations is large enough so
that the values of the natural logs of the slope are almost continuous.

From Eq.(\ref{genzdiff}), we have
\begin{eqnarray}
dn = \frac{dy}{ay^{z}+\epsilon},\label{dn}
\end{eqnarray}
so that
\begin{eqnarray}
\lambda(y_{in})=\frac{1}{n(y_{in})}\int\limits_{y_{in}}^{y_{out}}dy\frac{\ln|1
+azy^{z-1}|}{ay^{z}+\epsilon}.\label{lyexp2}
\end{eqnarray}
This gives the Lyapunov exponent for the system starting at $y_{in}$ and 
ending at $y_{out}$. 

In the logistic map or any other system, the entrance into the tube is random.
Since the starting points are randomly distributed, it is more useful to derive a
formula for the average Lyapunov exponent per pass. Using the above formula for 
$\lambda(y_{in})$, we can calculate the average value of the Lyapunov exponent and
obtain
\begin{eqnarray}
<\lambda>\equiv
\int\limits_{-y_{out}}^{y_{out}}dy_{in}\lambda(y_{in})P(y_{in}),\label{avgly}
\end{eqnarray}
where the probability function $P(y_{in})$ satisfies
\begin{eqnarray}
\int\limits_{-y_{out}}^{y_{out}}dy_{in}P(y_{in})=1.\label{prob}
\end{eqnarray}
For the present, let us consider the special case of a uniform distribution,
\begin{eqnarray}
P(y_{in})=\frac{1}{2y_{out}}.\label{uniform}
\end{eqnarray}
Using this probability distribution, we obtain
\begin{eqnarray}
<\lambda>=\frac{1}{2y_{out}}\int\limits_{-y_{out}}^{y_{out}}
dy_{in}\frac{I(y_{in})}{n(y_{in})} ,\label{avgly2}
\end{eqnarray}
where
\begin{eqnarray}
I(y_{in})\equiv \int\limits_{y_{in}}^{y_{out}}dy\frac{\ln|1
+azy^{z-1}|}{ay^{z}+\epsilon},\label{Idef}
\end{eqnarray}
and where $n(y_{in})$ is given by Eq.(\ref{n}) above. 

One approximation and a change of variables are necessary to 
make this formula more usable. One important step is to define the value 
for $y_{out}$ as
\begin{eqnarray}
y_{out}\equiv s\sqrt[z]{\frac{\epsilon}{a}},\label{scale}
\end{eqnarray}
where \lq\lq s\rq\rq \,is a positive scale factor that can be independently set
in  order to model a given system. Clearly, $s$ can not be arbitrarily large. A
natural requirement is that the derivative in (\ref{fdiff}) be positive, making
the possible \lq\lq throat\rq\rq \,of the channel end at the point where
$\frac{dF}{dy}=0$. This gives that
\begin{eqnarray}
s_{max} \equiv
z^{\frac{1}{1-z}}a^{\frac{1}{z(1-z)}}\epsilon^{-\frac{1}{z}},\label{smax}
\end{eqnarray}
is the maximum value of s for given $z$, $\epsilon$ and $a$. With the value of
$y_{out}$ from (\ref{scale}), the integral
$I(y_{in})$ can be simplified with a change of variables. Let
\begin{eqnarray}
y'=\frac{y}{y_{out}}.\label{vchange}
\end{eqnarray}
Therefore
\begin{eqnarray}
I(y_{in})=sa^{-\frac{1}{z}}\epsilon^{-1+\frac{1}{z}}
\int\limits_{\frac{y_{in}}{s\sqrt[z]{\frac{\epsilon}{a}}}}^{1}dy' \frac{\ln(1
+za^{\frac{1}{z}}\epsilon^{1-\frac{1}{z}}s^{z-1}y'^{z-1})}{s^{z}y'^{z}+1},
\label{I2}
\end{eqnarray}
where the absolute value in the natural log is no longer necessary. The Taylor
series expansion for natural log is ($|x|<1$)
\begin{eqnarray}
\ln
(1+x)=x-\frac{x^{2}}{2}+\frac{x^{3}}{3}-\frac{x^{4}}{4}+\ldots\, .\label{taylor}
\end{eqnarray}
Using this approximation we have simply
\begin{eqnarray}
I(y_{in})
\approx zs^{z}\int\limits_{\frac{y_{in}}{s\sqrt[z]{\frac{\epsilon}{a}}}}^{1}
dy' \frac{y'^{z-1}}{s^{z}y'^{z}+1}=\ln \left( \frac{s^{z}+1}{\frac{a}{\epsilon}
y_{in}^{z}+1}\right) ,\label{I3}
\end{eqnarray}
as long as 
\begin{eqnarray}
s << s_{max}. \label{approx}
\end{eqnarray}
Our simplified formula for the average Lyapunov exponent is now
\begin{eqnarray}
<\lambda>=\frac{1}{2s}a^{2/z}\epsilon^{1-2/z}
\int\limits_{-y_{out}}^{y_{out}}\frac{dy_{in}}
{\int\limits_{y_{in}\sqrt[z]{\frac{a}{\epsilon}}}^{s}\frac{d{\bar y}}{{\bar
y}^{z}+1}}\ln \left( \frac{s^{z}+1}
{\frac{a}{\epsilon} y_{in}^{z}+1}\right). \label{almost}
\end{eqnarray}
We now make the same scale change in the $y_{in}$ integral as in the $y$ integral
in (\ref{vchange}):
\begin{eqnarray}
{\hat y} \equiv \frac{y_{in}}{y_{out}}.\label{vchange2}
\end{eqnarray}
Therefore
\begin{eqnarray}
<\lambda>=\frac{1}{2}a^{1/z}\epsilon^{1-1/z}
\int\limits_{-1}^{1}\frac{d{\hat y}}{\int\limits_{s{\hat y}}^{s}\frac{d{\bar
y}}{{\bar y}^{z}+1}}\ln \left(
\frac{s^{z}+1}{s^{z}{\hat y}^{z}+1}\right). \label{final}
\end{eqnarray}
As one can see, for constant $s$ the average Lyapunov exponent varies as
$\epsilon^{1-1/z}$ with a constant of proportionality determined by the
parameters $a$ and $s$. In the case where $z=2$ this gives
\begin{eqnarray}
<\lambda>=\frac{1}{2}\sqrt{a\epsilon}
\int\limits_{-1}^{1}\frac{d{\hat y}}{\tan^{-1}(s)-\tan^{-1}(s{\hat y})}
\ln \left( \frac{s^{z}+1} {s^{z}{\hat y}^{z}+1}\right). \label{z=2}
\end{eqnarray}
For a general probability distribution, we would have instead
\begin{eqnarray}
<\lambda>=a^{1/z}\epsilon^{1-1/z}
\int\limits_{-1}^{1} d{\hat y} \frac{P({\hat y})}
{\int\limits_{s{\hat y}}^{s}\frac{d{\bar y}}{{\bar y}^{z}+1}}\ln \left(
\frac{s^{z}+1} {s^{z}{\hat y}^{z}+1} \right) d{\hat y},\label{final2}
\end{eqnarray}
where 
\begin{eqnarray}
\int_{-1}^{1} d{\hat y}P({\hat y}) = 1.\label{P}
\end{eqnarray}
In the case of constant scale factor $s$, we therefore see that the single
pass tangency channel Lyapunov exponent behaves like $\epsilon^{1-1/z}$.

\vspace{1cm}
\noindent
{\Large
{\bf 3. Tangency Channel Simulation}}
\vspace{.5cm}

The integral in Eq.(\ref{z=2}) was calculated using numerical methods and 
compared against numerical simulations of the logistic map ($z=2$). In all 
cases we used a simulation consisting of 10,000 Monte Carlo runs for each data
point. As can be seen in Figs.1-3, for values of $s=0.1$, 1.0 and 10, the
theoretical values agree with the simulation values for a uniform probability
distribution for small enough $\epsilon$. At low $\epsilon$, the agreement is
excellent, with the theoretical value straddled by the upper and lower error
values of the simulation. In the $s=0.1$ simulation, the assumption that the
discrete sum can be approximated by a continuous integral breaks down at large
enough $\epsilon$ due to the very small number of iterations in the channel. For
the $s=10$ simulation, the natural log approximation, Eq.(\ref{taylor}), starts to
break down and is the main cause of the divergence between theory and simulation.
The least divergence between theory and simulation occurs when $s\approx 1$ . The
calculations become more time consuming at larger s due to the increased number
of iterations necessary to pass through the channel.

We examined the more general expression, Eq.(\ref{final2}), when $z=2$
for other probability distributions including a Gaussian and a $|y|$ distribution
of normal deviates. Although these results are not illustrated here, the
theoretical and simulation results were again in excellent agreement for small
enough $\epsilon$. We also examined the $\epsilon$ dependence for higher
universality  classes. However, due to the large amount of computer time it takes to
do such simulations, we have data only for one additional $z$ value. For a
universality class of $z=4$, the Lyapunov dependence should be $\epsilon^{3/4}$,
which is clearly confirmed in Fig.~4.

The Section 2 expressions for $<\lambda>$ are displayed in a form appropriate for a
simulation at a fixed value of the scale, $s$. However, the approximation employed,
Eq.(\ref{approx}), is simply the condition that the channel length be much smaller
than the total gate size (determined by the tangency point and the point at
$\frac{dF}{dy}=0$). Thus, the expression Eq.(\ref{final2}) holds also for a
fixed tangency channel fraction, $f$,
\begin{eqnarray}
f\equiv \frac{y_{out}}{(az)^{\frac{1}{1-z}}}= s
\epsilon^{\frac{1}{z}}z^{\frac{1}{z-1}}a^{\frac{1}{z(z-1)}},\label{fagain}
\end{eqnarray}
as long as
\begin{eqnarray}
f<<1. \label{f}
\end{eqnarray}
The quantity $(az)^{\frac{1}{1-z}}$ is just the total model gate size. For $z=2$ the
relationship between
$f$ and $s$ is simply
\begin{eqnarray}
f = 2s\sqrt{a\epsilon}.\label{z=2f}
\end{eqnarray}

When the change from $s$ to $f$ is made in Eq.(\ref{final2}), the result is no
longer proportional to $\epsilon^{1/2}$ but in fact goes to a constant at small
values of $\epsilon$. Mathematically, this is due to the fact that the denominator,
proportional to the number of iterates in the channel for a starting position
${\hat y}$, falls off like $\epsilon^{1/2}$ when ${\hat y}\approx 1$. Physically, the
Lyapunov exponent is being dominated by the small number of iterates associated with
entrances on the far side of the narrow channel. Fig.~5 shows the
predicted and measured values of the Lyapunov exponent, $<\lambda>$ for the
case $f=0.1$, i.e., the channel length is one-tenth the size of the gate, when the
entrance probability is again uniform. As expected, and unlike the cases presented
above at fixed $s$, the value of the Lyapunov exponent becomes constant at small
$\epsilon$, the theoretical value remaining about $10\%$ larger than simulation.
This amount of deviation is what we would expect since the term kept in the
expansion of the natural log in Eq.(\ref{taylor}) is of order $f$ and the first
neglected term is of order $f^{2}$. Measurements at fixed $f$ will be important for
the simulations done in the full logistic map to be considered in the next Section.

\vspace{1cm}
\noindent
{\Large
{\bf 4. Logistic Map Simulation}}
\vspace{.5cm}

As was pointed out earlier, there are two manners in which Lyapunov exponents can
be defined in channel simulations, which we called the single pass and many pass
definitions. We have already described and illustrated a single pass simulation.
As an example of a many pass situation, we consider a simulation of the
tangency gates in the logistic map. In doing so, we find it convenient to first
describe the details and results of the numerical simulation. The theory and the
model used to fit to the data will then be developed together in Section 5.

As outlined in the Introduction, the equation for the third iterate
of the logistic map for small $\epsilon$ may be expanded in a Taylor series about
each of the tangency points. The positions of the tangency points are given to
high precision in Table I. (Knowledge of any one gives the others
through the basic recurrence relation.) Expanding to third order in
$(x-x_{c})$, one has for the third iterate, 
\begin{eqnarray}
F^{(3)}(x)=
x_{c}+(x-x_{c})+a_{c}(x-x_{c})^{2}+c_{c}(x-x_{c})^{3}+b_{c}(R_{c}-R).\label{fthing}
\end{eqnarray}
The values of $a_{c}$, $b_{c}$ and $c_{c}$ are given in Table I. Again introducing
$y_{n}=\frac{x_{n}-x_{c}}{b_{c}}$, the difference equation for all three
of the gates takes the form (note that $c_{c}b_{c}^{2}=-196$ to high accuracy for
all gates),
\begin{eqnarray}
y_{n+1}-y_{n}=\epsilon +ay_{n}^{2}-\frac{2}{49}a^{2}y_{n}^{3},\label{ndiff}
\end{eqnarray}
where the constant \lq\lq a \rq\rq takes on the value
\begin{eqnarray}
a=69.29646455628\ldots\,\, .
\end{eqnarray}

An interesting aspect of the simulation is the exclusion of certain $x$-values
from the logistic map at finite $\epsilon$. We have labeled the tangency gates in
increasing order of their $x_{c}$ values. Referring to the third iterate map, shown
for $\epsilon=0$ in Fig.~6, it is clear by drawing a horizontal line that
gate 1 is reachable only under steady-state conditions from points
close to point C in the Figure. Likewise, gate 3 is reachable only from previous
iterates starting on or near points A or B in the figure. (Since $x=0.5$
is a symmetry point in the mapping, the values of $F^{(3)}(x)$ at points
A and B are the same.) Note in this context that the laminate flow through gates 1
and 2 is from smaller $x$-values to larger ones, whereas the flow through gate 3 is
in the opposite direction. Iterates entering gate 1 from the point C will actually
enter at the value $x_{L}\equiv F^{(3)}(x_{C})$; points between $x=0$ and
$x=x_{L}$ will never be reached. This is after a possible transient of a single
iteration. Likewise, iterates entering gate 3 from points A and B will
enter at the value $x_{R}\equiv F^{(3)}(x_{A,B})$; points between $x=x_{R}$
and $x=1$ are never reached, again after a possible 1-iteration transient.
By drawing a horizontal line, gate 2 is seen to be reachable from 4 separate
$x$-regions (excluding the points to the left of $x_{L}$ and to the right of
$x_{R}$).

The measurements leading to values and variances of the Lyapunov exponents were taken
from a single trajectory of $1.6$ million iterations of the third-iterate logistic
map at each $\epsilon^{1/2}$ value following an initial \lq\lq heating\rq\rq to
remove the transient $x$-values. Monte Carlo error bars for the Lyapunov exponents
were then measured by breaking the single trajectory into 100 bins. Runs were made
both at fixed gate fraction, $f$, as well as at fixed scale factor, $s$.

Fig.~7 shows a ${\rm log}_{10}$ plot of the contribution of each of the three gates
when the fraction of the gate being measured is $f=0.1$, the same used in Fig.~5.
Note that $f=0.1$ indicates the fraction of the model gate size, characterized
by Eq.~(\ref{what}), not the actual gate size. The ratio of the model to actual
gate length is about $1.025$. The gate Lyapunov exponents, $<\lambda >_{g}$, are
normalized to the number of third-iterate hits within each gate, which at small
$\epsilon$ just approaches $1/3$ of the total iterations. We notice that even at
the larger $\epsilon$ values the contribution from different gates is the same within
errors.

Fig.~8 shows the contribution of each of the three gates to the Lyapunov exponent,
$<\lambda>_{g}$, in a simulation with a fixed value of the scale factor, $s=1.0$. 
Eq.(\ref{it}) implies that such a simulation will have a mixture of half
iterations inside the gates and half outside, as indeed is observed. The
error bars here are larger in a relative sense than in Fig.~7 because the gate size
is shrinking like $\epsilon^{1/2}$ (see Eq.(\ref{scale})), leading to smaller
statistics. Also unlike Fig.~7 there are significantly different contributions from
the three gates at larger $\epsilon$ values, the middle gate having an enhanced
$<\lambda >_{g}$. It is only at values of $\epsilon$ below and including
$\epsilon^{1/2}=0.512\times 10^{-3}$ that a distinction between the gates can no
longer be seen. However, this may simply be the result of the larger statistical
fluctuations present at smaller $\epsilon$.

Note that the fixed $f$ data in Fig.~7 shows an approximate $\epsilon^{1/2}$
behavior, while the fixed $s$ simulation in Fig.~8 behaves approximately as
$\epsilon$ at small $\epsilon$ values. These behaviors are in contrast to the
simulations in Section 3 where the fixed $f$ data went to
a constant at small $\epsilon$ and the fixed $s$ data behaved like $\epsilon^{1/2}$.
The $\epsilon$ behavior of the logistic map Lyapunov exponents will be commented on
further in the next Section.

In order to model the Lyapunov exponents for the gates, it is necessary to have an
understanding of the entrance probability for the gates as a function of position
in the gate. There are two ways in which a laminate flow may begin in the tangency
channel. Primarily, entrance into the gate occurs as a continuous flow from
just outside the gate. Alternatively, the flow can begin in a discontinuous
fashion from a disjoint region of the map. These two entrance routes will be
termed the {\it continuous} and {\it discontinuous} types, respectively. Fig.~9
presents a measurement of the binned discontinuous entrance rate, $n_{d}$, for the
first gate of the $f=0.1$ simulation at $\epsilon^{1/2}=0.128\times 10^{-3}$. The
data is divided into 19 bins with bin size of $\Delta  x_{bin}=0.5624 \times
10^{-4}$, which is just the model gate width divided by 100. The first bin (which
would have extended from -10 to -9 in the figure units) contains both continuous and
discontinuous entrances. Since we have not attempted to separate the discontinuous
from the continuous entrances in this bin and because the total rate is
off scale, this first bin is not shown. The entrance rate seems to be fairly uniform
in this Figure; gate 2 and (the mirror image of) gate 3 look very similar.
This approximate entrance uniformity for small $f$ will be useful in setting up a
simple model of the gate contribution to the Lyapunov exponent, which will be
described in the next Section.

\vspace{1cm}
\noindent
{\Large
{\bf 5. Simple Model}}
\vspace{.5cm}

We will now develop theoretical expressions for the $\epsilon$-dependence of the
Lyapunov exponents for the logistic map. For this purpose we need to develop
expressions for $<n>$ and $<\lambda >$ in a many pass simulation, as opposed to the
single pass considerations in Sections 2 and 3. In a many pass simulation, the
number of iterations in the gate will be weighted by the entrance {\it rate} rather
than probability. Therefore, Eq.(\ref{it}), generalized to an arbitrary probability
distribution, is replaced with
\begin{eqnarray}
<n>_{g}= \frac{1}{2}\int_{-1}^{1}d{\hat y}\,\frac{dN({\hat y})}{d{\hat y}}n(y_{in}),
\label{huh}
\end{eqnarray}
where
\begin{eqnarray}
n(y_{in})=\frac{1}{\sqrt{a\epsilon}}(\tan^{-1}(\frac{f}{2\sqrt{a\epsilon}})-\tan^{-1}
(\frac{f{\hat y}}{2\sqrt{a\epsilon}})).
\label{nn}
\end{eqnarray}
We also are using the dimensionless variable ${\hat y}$ introduced in
Eq.(\ref{vchange2}) ($y_{in}={\hat y}y_{out}$). The functional form of
$\frac{dN({\hat y})}{d{\hat y}}$ has yet to be specified. Note that $n_{d}$ in
Fig.~9 is given by $n_{d}=\frac{dN({\hat y})}{d{\hat y}}\Delta {\hat y}_{bin}$ where
in this case $\Delta {\hat y}_{bin}=0.1$.

The form for the Lyapunov exponent, Eqs.(\ref{avgly2}) and (\ref{Idef}), must also
be modified. It is again the rate rather than the probability which is relevant. In
addition, for a many pass simulation the Lyapunov
integrand in Eq.(\ref{avgly2}) must be weighted by the ratio of the number of
iterations for each passage through the channel, $n(y_{in})$, to the average total
number of iterations, $<n>_{g}$, resulting in
\begin{eqnarray}
<\lambda>_{g}= \frac{1}{2<n>_{g}}\int_{-1}^{1} d{\hat y}\,\frac{dN({\hat
y})}{d{\hat y}}I(y_{in}), \label{blaa}
\end{eqnarray}
where
\begin{eqnarray}
I(y_{in})=\frac{2}{f}\int_{{\hat y}}^{1} dy'\,\frac{\ln
(1+fy')}{y'^{2} +\frac{4a\epsilon}{f^{2}}}.\label{Iguy}
\end{eqnarray}
Making the same approximation as in Eq.(\ref{taylor}) above, this simplifies to
\begin{eqnarray}
<\lambda>_{g}= \frac{1}{2<n>_{g}}\int_{-1}^{1}d{\hat y}\,\frac{dN({\hat y})}{d{\hat
y}}\ln \left( \frac{1+\frac{4a\epsilon}{f^{2}}} {{\hat
y}^{2}+\frac{4a\epsilon}{f^{2}}} \right).
\label{lambda}
\end{eqnarray}
We use $f$ instead of $s$ in these formulas since most of the simulations in
the following will use fixed $f$.

As we have seen in the last Section, the
Lyapunov exponents from the three gates are apparently indistinguishable at small
enough $\epsilon$. A very useful simplification therefore is to ignore the
distinction between the gates and model the Lyapunov exponent as if there were only
two regions, the gate (or periodic) region and the outside (or chaotic) region. In
addition, Fig.~9 suggests that a reasonable model of the channel region is to assume
that the discontinuous entrance rate is uniform. This last simplification is only
possible for small enough $f$, the entrance rate in the complete tangency channels
being far from uniform. These assumptions will allow us to construct a very simple
model of the $\epsilon$ dependence of the Lyapunov exponents. The choice of $f=0.1$
to separate the two regions is arbitrary. A smaller value would result in an even
more uniform entrance rate than Fig.~9; however, one would also loose statistics
because of the smaller gate size. As we will see, the gate fraction $f$ will be used
to formally separate the outside and inside gate Lyapunov exponent behaviors.

A new aspect of modeling the actual gates of the logistic map is the fact that
the entrance to the gates can be from flow further up the channel or from
a completely disjoint part of the map. These possibilities were termed the continuous
and discontinuous routes in the last Section. One may show that the continuous
entrances occur within a scaled distance of $\sim f/2 +2a\epsilon/f$ from the
lower limit ($-1$) of the integrals in Eqs.(\ref{huh}) and (\ref{blaa}). In
Fig.~9 this contribution would extend about halfway through the first (deleted) bin.
Since these entrances always occur in a narrow range of the integrations in
these equations for the range of $\epsilon$ considered, it is reasonable to model
this contribution by a Dirac delta function located at the lower limit, ${\hat
y}=-1$. In addition, we saw in Fig.~9 that the discontinuous part of the entrance
rate was approximately uniform. Thus we will model the entrance rate with
\begin{eqnarray}
\frac{dN({\hat y})}{d{\hat y}}=2N_{c}\delta({\hat y}+1) + \frac{dN_{d}}{d{\hat y}},
\label{Neq}
\end{eqnarray}
where $N_{c}$ and $\frac{dN_{d}}{d{\hat y}}$ are constants in ${\hat y}$.
This gives a simple model of the contributions to $<n>_{g}$ and $<\lambda >_{g}$ from
the logistic gates:
\begin{eqnarray}
<n>_{g}= \frac{1}{\sqrt{a\epsilon}}\tan^{-1}(\frac{f}{2\sqrt{a\epsilon}})\left(
2N_{c}+\frac{dN_{d}}{d{\hat y}}
\right), \label{nnn}
\end{eqnarray}
\begin{eqnarray}
<\lambda>_{g}= \frac{1}{2<n>_{g}}\frac{dN_{d}}{d{\hat y}}\int_{-1}^{1}d{\hat y}\ln
\left( \frac{1+\frac{4a\epsilon}{f^{2}}} {{\hat y}^{2}+\frac{4a\epsilon}{f^{2}}}
\right).
\label{lambdax}
\end{eqnarray}
Notice that $N_{c}$ drops out of the expression for $<\lambda >_{g}$. In fitting the
data, we also need the expression for the small $\epsilon$ limit of Eq.(\ref{nnn}),
which we will call $n_{T}$:
\begin{eqnarray}
n_{T}\equiv \frac{\pi}{2\sqrt{a\epsilon}}\left( 2N_{c}+\frac{dN_{d}}{d{\hat y}}
\right).
\label{nT}
\end{eqnarray}

As pointed out previously, $f$ represents a separation parameter for the gate and
outside regions. Inside the gates one expects from the previous results
that the number of iterations associated with a given traverse of the gate
will increase like $\epsilon^{-1/2}$ at small $\epsilon$, independent of the
form of the entrance probability. Thus in the many pass simulations of the full
logistic map, one expects any given fraction $f$ of the gates at small $\epsilon$ to
eventually contain essentially all iterates. This means from Eq.(\ref{nT}) that the
quantity $2N_{c}+\frac{dN_{d}}{d{\hat y}}$ must scale like $\epsilon^{1/2}$ for
$n_{T}$ to become constant. This behavior will be assumed for these quantities
individually. In addition, we assume that for small enough gate fraction the
continuous entrance rate is uniform. Given these assumptions, the parameters $N_{c}$
and $\frac{dN_{d}}{d{\hat y}}$ can be parameterized as,
\begin{eqnarray}
N_{c} &\equiv& \sqrt{a\epsilon}(1-f) n_{T} K_{c}, \\ \label{K1}
\frac{dN_{d}}{d{\hat y}}&\equiv&\sqrt{a\epsilon}f n_{T} K_{d}, \label{K2}
\end{eqnarray}
where $K_{c}$ and $K_{d}$ are constants. 

We can now better understand the $\epsilon$ dependencies in the Lyapunov
exponents in the logistic map simulation seen in the last Section. For a many
pass simulation we expect the model Lyapunov exponent from Eqs.(\ref{lambdax})
and (\ref{K2}) for fixed $f$ to behave like $\epsilon^{1/2}$ at small $\epsilon$.
This is because $<n>_{g}$ saturates to the value $n_{T}$ while the rate in
Eq.(\ref{K2}) goes like $\epsilon^{1/2}$. In contrast, for fixed scale $s$ the
exponent now acquires an extra $\epsilon^{1/2}$ factor from the gate fraction $f$ in
(\ref{K2}) and is expected to decrease like $\epsilon$, as
was seen in the actual simulations. This is just a result of
the shrinking gate size of the fixed $s$ simulation.

As a result of the number flow into the gate regions as
$\epsilon$ decreases, the outside region becomes sparsely visited, but with the same
local density of iterates. (The shape of the outside region changes very little for
small $\epsilon$.) This implies the Lyapunov exponent measured in the outside region
will go to a constant for small $\epsilon$. The flows being described are all due to
the result Eq.(\ref{n}) for $n(y_{in})$ and can be thought of as
applications of the renormalization flow arguments, without external noise, 
in [J.\ E.\ Hirsch {\it et. al.}, 1982] and [ B.\ Hu and J.\ Rudnick,
1982].

Although the emphasis here is on the gate Lyapunov exponents, one can now
make a rough model of the complete logistic map exponent.
Letting $<\lambda>_{g}$ represent the Lyapunov exponent expression for the
fixed $f$ gate (laminar region) from Eq.(\ref{lambdax}) and $<\lambda>_{o}$ be the
outside (chaotic region) contribution, the expression for the Lyapunov exponent
for the complete logistic map in this model is
\begin{eqnarray}
<\lambda >^{3rd} =
\frac{n_{o}}{n_{T}}<\lambda>_{o}+\frac{n_{g}}{n_{T}}<\lambda>_{g},
\label{fullL}
\end{eqnarray}
where $n_{o}+n_{g}=n_{T}$. That is, $<\lambda >^{3rd}$ is just assumed to be a sum of
the exponents $<\lambda>_{g}$ and $<\lambda>_{o}$ weighted by the relative number of
iterations spent in the two regions. A more fundamental description of the logistic
map would independently calculate $<\lambda>_{o}$. However, from the previous
arguments we expect $<\lambda>_{o}$ to simply be a constant at small $\epsilon$. It
will be evaluated from a fit to the data. 

Numerically, the gate contribution in Eq.(\ref{fullL}) is only about $1\%$ of the
total through most of the $\epsilon$ range considered for $f=0.1$. Both terms in
Eq.(\ref{fullL}) go like $\epsilon^{1/2}$, but in different ways. The outside term
behaves like $\epsilon^{1/2}$ because the outside number $n_{o}$ has this dependence;
the gate term also behaves this way because $<\lambda>_{g}$ itself goes like
$\epsilon^{1/2}$ at fixed $f$ as explained above. Note that all Lyapunov exponents in
Eq.(\ref{fullL}) are normalized to the number of third-iterate steps in the
simulation. This is symbolized by writing $<\lambda>^{3rd}$ for the complete
logistic map Lyapunov exponent. We must remember to divide this value by three to
calculate the usual single-iterate value:
\begin{eqnarray}
<\lambda>^{1st}=\frac{1}{3}<\lambda>^{3rd}. \label{twolams}
\end{eqnarray}

Fig.~10 presents the results of fitting Eq.(\ref{lambdax}) to the data for
the $f=0.1$ gate exponent and Fig.~11 gives the measured and
model $<\lambda>^{1st}$ values for the complete logistic map. There are
three parameters needed to do these fits: $K_{c}$, $K_{d}$ and $<\lambda>_{o}$. To
evalute these constants we simply fit the values of these expressions to the
measured values of $<\lambda>^{1st}$ and $<\lambda>_{g}$ at a value of
$\epsilon^{1/2}=0.128\times 10^{-3}$, near the middle of the exponential range in
these figures, as well as the maximum number of gate iterations, $n_{T}$. Since
we are averaging over the properties of all three gates, $n_{T}=
\frac{1}{3}\times 1.6 \times 10^{6}$ for the simulation in Section 4. This gives
$K_{c}=0.358\times 10^{-2}$, $K_{d}=0.739\times 10^{-2}$ and $<\lambda>_{o}=0.962$. 

Of course since Figs.~10 and 11 were used to fit
the model parameters, we need an independent test of how well the model truly
represents the data. For this purpose we also present Fig.~12 and Table II. Fig.~12
compares the model results against the data for a $s=1.0$
simulation. As explained above, this data falls like $\epsilon$. The theoretical
results are satisfactory although seem somewhat high compared to measurement. In
addition in Table II we give the fit results for the rates $N_{c}$ and
$\frac{dN_{d}}{d{\hat y}}$ compared to measurement. The measured value for
$\frac{dN_{d}}{d{\hat y}}$ is actually just an average over all three gates of
binned data similar to Fig.~9, and the value for $N_{c}$ is the average value in the
three entrance bins minus the average of the entrances in the other bins. The results
for $N_{c}$ are good but the fit values of $\frac{dN_{d}}{d{\hat y}}$ are
approximately a factor of 2 larger than measurement. As we will see in the next
Section, this is largely the result of an inaccurate characterization of the inverse
number density,
$\frac{dy}{dn}$.

\vspace{1cm}
\noindent
{\Large
{\bf 6. Improved Model}}
\vspace{.5cm}

Although the expression for the inverse number density, Eq.(\ref{genzdiff}), is
symmetric (even) in $y$ for $z=2$, there are at least three sources of asymmetry in
the actual gate. First, the discontinuous entrance rate $\frac{dN_{d}}{d{\hat y}}$
raises the value of the hit density on the exit sides of all three gates. Second,
the term proportional to $(x-x_{c})^{3}$ in Eq.(\ref{fthing}) shows that there is a
small intrinsic asymmetry in the shape of the gates themselves, raising the hit
density in the same sense. Most interestingly, there is also a contribution due to
the finite step size of the laminar flow through the gates which contributes even
for a perfectly symmetric gate. This will now be described.

Remembering that a finite difference gave rise to the left hand side of
Eq.(\ref{genzdiff}), a more accurate differential characterization of the iterative
flow is
\begin{eqnarray}
\frac{dy}{dn}+\frac{1}{2}\frac{d^{2}y}{dn^{2}}=ay^{2}+\epsilon.\label{diff2}
\end{eqnarray}
We will use the method of successive approximants to evaluate the second derivative
term. To zeroth order,
\begin{eqnarray}
\frac{dy}{dn}|_{0}=ay^{2}+\epsilon.\label{diff3}
\end{eqnarray}
Thus the lowest order result for the second derivative is just
\begin{eqnarray}
\frac{d^{2}y}{dn^{2}}|_{0}=2ay(ay^{2}+\epsilon).\label{diff4}
\end{eqnarray}
Inserting this back into the starting point, Eq.(\ref{diff2}), we now have the
improved result
\begin{eqnarray}
\frac{dy}{dn}|_{1}=ay^{2}+\epsilon (1-ay)-a^{2}y^{3}.\label{diff5}
\end{eqnarray}
With this improvement, a better formula for the gate Lyapunov exponent is given
by
\begin{eqnarray}
<\lambda >_{g} & = & \frac{2N_{c}}{<n>_{g}}\int_{-1}^{1} dy'
\frac{y'}{y'^{2}+\frac{4a\epsilon}{f^{2}}(1-\frac{f}{2}y')
-\frac{f}{2}y'^{3}} \\
\nonumber & + & \frac{dN_{d}}{d{\hat y}}\frac{1}{<n>_{g}}\int_{-1}^{1}d{\hat
y}\int_{{\hat y}}^{1}
dy'\frac{y'}{y'^{2}+\frac{4a\epsilon}{f^{2}}(1-\frac{f}{2}y')-\frac{f}{2}y'^{3}}.
\label{newlam}
\end{eqnarray}
Note that the intrinsic contribution to the inverse hit rate from
Eq.(\ref{ndiff}) is of the same sign but 49/2 times smaller than the term from
Eq.(\ref{diff5}) and so is neglected. (The intrinsic term would also slightly alter
the numerator; see Eq.(\ref{lydef}).)

Notice that the continuous entrance term, $N_{c}$, now {\it does} contribute to
the expression for $<\lambda>_{g}$ since the inverse number density is no longer an
even function. Unfortunately, the innner integral can no longer be done
analytically and a double integral survives. Because the emphasis here is on
modeling the Lyapunov exponent and because of the difficulty of performing the
numerical integration leading to $n_{T}$ at small values of $\epsilon$, we have not
attempted to make the same correction in the expression for $<n>_{g}$. Thus, we
continue to use the expression Eq.(\ref{nnn}) above for $<n>_{g}$ in the gate
region. 

When the same sort of fit is made to the simulation data as in Section 5, there is
surprisingly little change in the functional forms in Figs.~10, 11 and 12, although
the inverse number rates of the two models are considerably different and the
$N_{c}$ term is now contributing about $50\%$ of the total. The new fit gives
$K_{c}\simeq K_{d} = 0.379\times 10^{-2}$. (The value of $<\lambda>_{o}$ does not
change from the previous fit.) The major improvement occurs in the value of the
discontinuous entrance density,
$\frac{dN_{d}}{d{\hat y}}$ (see the \lq\lq Improved model\rq\rq columns of Table
II), which is now within $5\%$ of the measured value at $\epsilon^{1/2}=0.128\times
10^{-3}$, where the fit is actually made. However, the value for $N_{c}$ has
increased and is now approximately $6\%$ large compared to the measured value. (Note
that $2N_{c}+\frac{dN_{d}}{d{\hat y}}$ in Table II is required to have the same
value in the two models because the form for
$<n>_{g}$ is unchanged.) This problem should be cured when the more accurate result
for the number density implied by Eq.(\ref{diff5}) is used in the expression for
$n(y_{in})$ in Eq.(\ref{n}).

\vspace{1cm}
\noindent
{\Large
{\bf 7. Summary and Conclusions}}
\vspace{.5cm}

Tangent bifurcation or intermittent chaos is a common occurrence in 
systems that exhibit chaotic behavior. In these systems the intermittent
behavior can be modeled by differential and difference equations of some
universality class. We found that the Lyapunov exponent for isolated gates at
single channel pass can be modeled given the universality class, the parameters of
the difference equation, the scale factor $s$ or fraction $f$ of the gate size, and
the closeness factor $\epsilon$. Our main theoretical result for these systems,
subject to the restriction of a sufficient number of steps in the channel and the
small gate approximation in Eqs.(\ref{approx}) or (\ref{f}), is that the average
Lyapunov exponent is given by Eqs.(\ref{final2}) and (\ref{P}). Single pass numerical
simulations were consistent with these expressions. At
fixed scale factor $s$ these results gave a Lyapunov exponent proportional
to $\epsilon^{1-1/z}$ for a tangency channel with
general universality class, $z$. We also showed that a simulation at fixed gate
fraction $f$ gave a result which instead became constant at small
values of $\epsilon$ for $z=2$ due to a small number of entrances on the far side of
the narrow channel.

Simulations were also performed on the full logistic map near the
intermittent transition at $R=1+\sqrt{8}$. Modified expressions for the gate number,
Eq.(\ref{huh}), and Lyapunov exponent, Eq.(\ref{lambda}), for a many
channel pass simulation were motivated. A new aspect encountered in the
description of the actual tangency channels was the existence of a continuous flow
contribution into the tangency gate. Two phenomenological models of the channel were
constructed and examined. A very simple model was considered which was able to give
a fairly realistic characterization of the various Lyapunov exponents and the
continuous, $N_{c}$, and discontinuous, $\frac{dN_{d}}{d{\hat y}}$, entrance
parameters. We also derived a first-order correction to the inverse hit rate due to
the discrete nature of the iterative flow, which mainly improved the comparison with
the measured discontinuous entrance parameter.

Besides applying the finite discretization correction to the number density
expression, there is considerable room for improving the present model of the
logistic gates. For example, the modeling of the continuous
contribution to the gates as a delta function is clearly oversimplified; no attempt
has made no attempt to resolve the shape of the continuous entrance rate in the
binning procedure of Fig.~9. In addition, the approximation used for the logarithm,
Eq.(\ref{taylor}), can be removed at the cost of a more complicated numerical
evaluation. Finally, further discretization corrections to both
$<\lambda>_{g}$ and $<n>_{g}$ should result in an improved characterization of the
inverse number density, leading to better comparison with the measured rates and
functional behaviors at larger $\epsilon$ values.

\vspace{1cm}
\noindent
{\Large
{\bf Acknowledgments}}

\vspace{.5cm}

This work was supported in part by NSF Grants PHY-9424124 and PHY-9722073. Some of
the numerical calculations were performed on the Power Challenge Array at the
National Center for Supercomputing Applications.

\newpage
{\Large
{\bf References}}
\vspace{.5cm}
\begin{description}

\item \quad  Berg\'e, P., Dubois, M., Manneville, P., \& Pomeau, Y. [1980],
\lq\lq Intermittency in Rayleigh-B\'enard Convection\rq\rq, {\it J.\ Physique
Lett.} {\bf 41}, L341-L354.

\item \quad  Hirsch, J. E., Hubermann, B. A. \& Scalapino, D. J. [1982]
\lq\lq Theory of Intermittency\rq\rq, {\it Phys.\ Rev.} {\bf A25}, 519-532.

\item \quad Hirsch, J.E. \& Nauenberg, M. [1982], \lq\lq Intermittency in
the Presence of Noise: A Renormalization Group Formulation\rq\rq, {\it Phys.\ Lett.}
{\bf 87A}, 391-393.

\item \quad Hu, B. \& Rudnick, J. [1982], \lq\lq Exact Solutions to the
Feigenbaum Renormalization-Group Equations for Intermittency\rq\rq, {\it Phys.\
Rev.\ Lett.} {\bf 48}, 1645-1648.

\item \quad Jeffries, C. \& P\'erez, J. [1982], \lq\lq Observation of a
Pomeau-Manneville Intermittent Route to Chaos in a Nonlinear Oscillator\rq\rq,
{\it Phys.\ Rev.} {\bf A26}, 2117-2122.

\item \quad Manneville, P. \& Pomeau, Y. [1979] \lq\lq Intermittency and the
Lorenz Model\rq\rq, {\it Phys.\ Lett.} {\bf 75A}, 1-2. 

\item \quad Pomeau, Y. \& Manneville, P. [1980] \lq\lq Intermittent
Transition to Turbulence in Dissipative Dynamical Systems\rq\rq, {\it Commun.\
Math.\ Phys.} {\bf 74}, 189-197.

\item \quad Pomeau, Y., Roux, J. C., Rossi, A., Bachelart, S. \&
Vidal, C. [1981], \lq\lq Intermittent Behavior in the Belousov-Zhabotinsky
Reaction\rq\rq,  {\it J.\ Physique Lett.} {\bf 41}, L271-L273.
 
\item \quad Scheck, F. A. [1994] {\it Mechanics} (Springer, Berlin) 2nd ed.,
p.\ 371.

\item \quad Schuster, H. G. [1995] {\it Deterministic Chaos} (VCH, Weinheim)
3rd ed., pp.\ 79-102.

\item \quad Weh, W. J., \& Kao, Y. H. [1983], \lq\lq Intermittency in
Josephson Junctions\rq\rq, {\it Appl.\ Phys.\ Lett.} {\bf 42}, 299-301.

\end{description}

\newpage
\begin{table}
\caption{Values of the tangency points, $x_{c}$, and the
constants $a_{c}$, $b_{c}$ and $c_{c}$ in Eq.(39).}

\begin{tabular}{c|c|c|c|c} 

 & $x_{c}\quad$ & $a_{c}\quad$ & $b_{c}\quad$ & $c_{c}\quad$ \\ \hline

{\rm gate\, 1} & $0.1599288184463\dots$ & $88.91012989368\dots$ &
$0.7793989800616\dots$ & $-322.6535182739\dots$ \\ 
{\rm gate\, 2} & $0.5143552770620\dots$ & $34.14530797001\dots$ &
$2.029457886780\dots$ & $-47.58783903539\dots$ \\ 
{\rm gate\, 3} & $0.9563178419736\dots$ & $-310.6483669763\dots$ &
$-0.2230704292148\dots$ & $-3938.873792041\dots$

\end{tabular}
\end{table}

\begin{table}
\caption{Results from two models for the continuum number
contribution, $N_{c}$, and the discontinuous number density, 
$\frac{dn}{d{\hat y}}$ when $f=0.1$.}

\begin{tabular}{c|c|c|c|c|c|c}

&\multicolumn{2}{c|}{Simulation}
&\multicolumn{2}{c|}{Simple model}
&\multicolumn{2}{c}{Improved model}
\\ \hline

$\epsilon^{1/2}$ 
&\multicolumn{1}{c|}{$N_{c}$} 
&\multicolumn{1}{c|}{$\frac{dN_{d}}{d{\hat y}}$} 
&\multicolumn{1}{c|}{$N_{c}$}
&\multicolumn{1}{c|}{$\frac{dN_{d}}{d{\hat y}}$}
&\multicolumn{1}{c|}{$N_{c}$}
&\multicolumn{1}{c}{$\frac{dN_{d}}{d{\hat y}}$}

\\ \hline

$0.8\times 10^{-5}$ & $0.1021(6)\times 10^{2}$ & $0.112(5)\times 10^{1}$ 
& $0.101\times 10^{2}$ & $0.232\times 10^{1}$ & $0.107\times 10^{1}$ & $0.119\times
10^{1}$ \\

$0.16\times 10^{-4}$ & $0.2028(8)\times 10^{2}$ & $0.224(6)\times 10^{1}$
& $0.203\times 10^{2}$ & $0.464\times 10^{1}$ & $0.214\times 10^{2}$ & $0.237\times
10^{1}$ \\

$0.32\times 10^{-4}$ & $0.403(1)\times 10^{2}$ & $0.466(8)\times 10^{1}$
& $0.406\times 10^{2}$ & $0.927\times 10^{1}$ & $0.429\times 10^{2}$ & $0.474\times
10^{1}$ \\

$0.64\times 10^{-4}$ & $0.804(2)\times 10^{2}$ & $0.93(2)\times 10^{1}$
& $0.812\times 10^{2}$ & $0.185\times 10^{2}$ & $0.857\times 10^{2}$ & $0.949\times
10^{1}$ \\

$0.128\times 10^{-3}$ & $0.1607(2)\times 10^{3}$ & $0.181(2)\times 10^{2}$
& $0.162\times 10^{3}$ & $0.371\times 10^{2}$ & $0.171\times 10^{3}$ &$0.190\times
10^{2}$ \\

$0.256\times 10^{-3}$ & $0.3181(3)\times 10^{3}$ & $0.368(2)\times 10^{2}$
& $0.325\times 10^{3}$ & $0.742\times 10^{2}$ & $0.343\times 10^{3}$ & $0.380\times
10^{2}$ \\

$0.512\times 10^{-3}$ & $0.6274(4)\times 10^{3}$ & $0.718(4)\times 10^{2}$
& $0.649\times 10^{3}$ & $0.148\times 10^{3}$ & $0.686\times 10^{3}$ & $0.759\times
10^{2}$ \\

$0.1024\times 10^{-2}$ & $0.1216(1)\times 10^{4}$ & $0.1411(5)\times 10^{3}$
& $0.130\times 10^{4}$ & $0.297\times 10^{3}$ & $0.137\times 10^{4}$ & $0.152\times
10^{3}$ \\

$0.2048\times 10^{-2}$ & $0.2297(1)\times 10^{4}$ & $0.2623(7)\times 10^{3}$
& $0.260\times 10^{4}$ & $0.594\times 10^{3}$ & $0.274\times 10^{4}$ & $0.304\times
10^{3}$  \\

$0.4096\times 10^{-2}$ & $0.4135(2)\times 10^{4}$ & $0.4769(9)\times 10^{3}$
& $0.519\times 10^{4}$ & $0.119\times 10^{4}$ & $0.548\times 10^{4}$ & $0.607\times
10^{3}$

\end{tabular}
\end{table}
\newpage
{\Large
{\bf Figure Captions} }
\vspace{.5cm}
     
\begin{enumerate}

\item  Simulation of the system described by Eq.(\ref{genz}) compared to the
Lyapunov exponent given by Eq.(\ref{z=2}) ($z=2$; uniform entrance probability).
We plot $\log_{10} <\lambda>$ against $\log_{10} \epsilon^{1/2}$; the
prediction (\ref{z=2}) is given by the dotted line. The Monte Carlo error bars on
the calculation are extremely small and are given by the data point
bars. We are using $a=34$ (the same as in [Hirsch {\it et.\ al.}, 1982]) and
$s=0.1$, with $\epsilon^{1/2}$ ranging in value from $0.8\times 10^{-5}$ upwards
by factors of 2. 
 
\item  The same as Fig.~1 but for s=1.  

\item  The same as in Fig.~1 except for $s=10$. Note that the largest
$\epsilon^{1/2}$ value, present in Figs.~1 and 2, violates the
bound $s<s_{max}$ of the text and has been excluded.

\item  The case of $z=4$, $s=0.1$ $a=34$ and uniform entrance probability.

\item Contribution to the Lyapunov exponent, as a function of $\log_{10}
\epsilon^{1/2}$, from a simulation involving 10,000 gate entrances when $f$, the
gate fraction, is set to $0.1$, the entrance probability
is uniform and $a=34$. The theoretical result from Eq.(\ref{z=2}) is given by the
dotted line.

\item The third iterate of the logistic equation, $F^{(3)}(x)$, as a function
of $x$ when $\epsilon=0$. The three points where the third iterate makes
tangential contact with the $45$-degree line are the tangency points. The points A, B
and C in the map are discussed in the text.

\item Contributions of the three tangency gates to the logistic map Lyapunov
exponent, $<\lambda >_{g}$ for $f=0.1$. Gate 1 data is given by the circles, gate 2
by the squares and gate 3 by the triangles. Note that the ordinate values of the
gate 1 and 3 data points have been shifted to the left and right, respectively, for
clarity of presentation.

\item Contributions of the three tangency gates to the logistic map Lyapunov
exponent, $<\lambda >_{g}$ for $s=1.0$. The meaning of the symbols is the same as in 
Fig.~7.

\item The number of discontinuous entrances, $n_{d}$, for gate 1 when $f=0.1$
and $\epsilon^{1/2}=0.128\times 10^{-3}$ as a function of bin number centered about
the first tangency point, $x_{c}$. Each data point above corresponds to entrances in
a bin size of $\Delta x_{bin}= 0.5624\times 10^{-4}$. (The bin extending
from -10 to -9 is not shown; see the text.)

\item Comparison of the model results for the Lyapunov exponent for the $f=0.1$
simulation with the averaged data from Fig.~7. The model values, given by a dotted
line (simple model) and a solid line (improved model), are indistinguishable. The
data point at $\log_{10}(0.128\times 10^{-3})=-3.893\ldots$ is used for the fit.

\item Comparison of the model results for the Lyapunov exponent ($1st$ iterate) for
the entire map with the data. The model values, given by a dotted line (simple model)
and a solid line (improved model), are again indistinguishable. The data point at
$\log_{10}(0.128\times 10^{-3})=-3.893\ldots$ is used for the fit.

\item Comparison of the model results for the Lyapunov exponent for the $s=1.0$
simulation with the averaged gate data from Fig.~8. The model values are given by a
dotted line (simple model) and a solid line (improved model).

\end{enumerate}

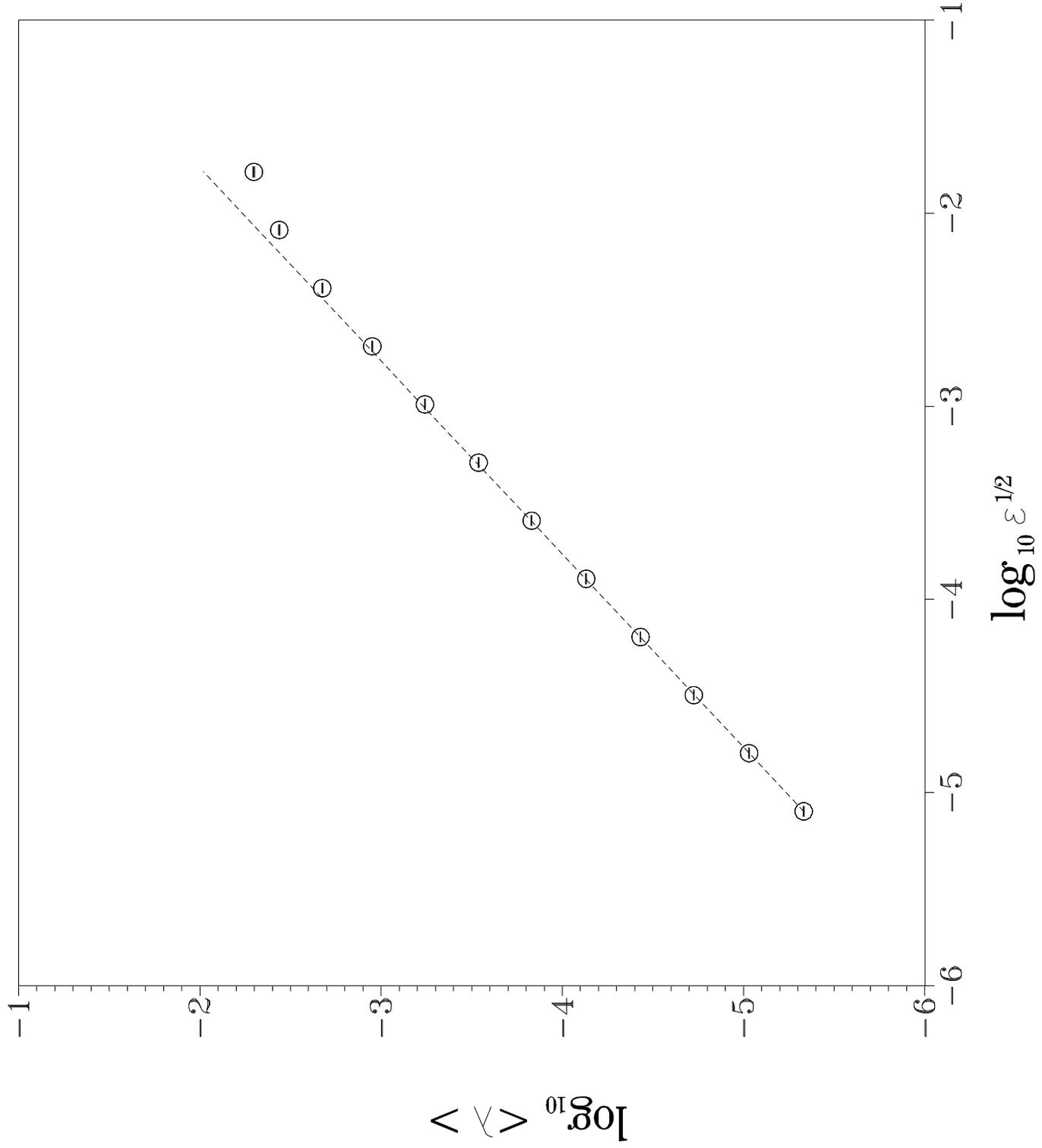
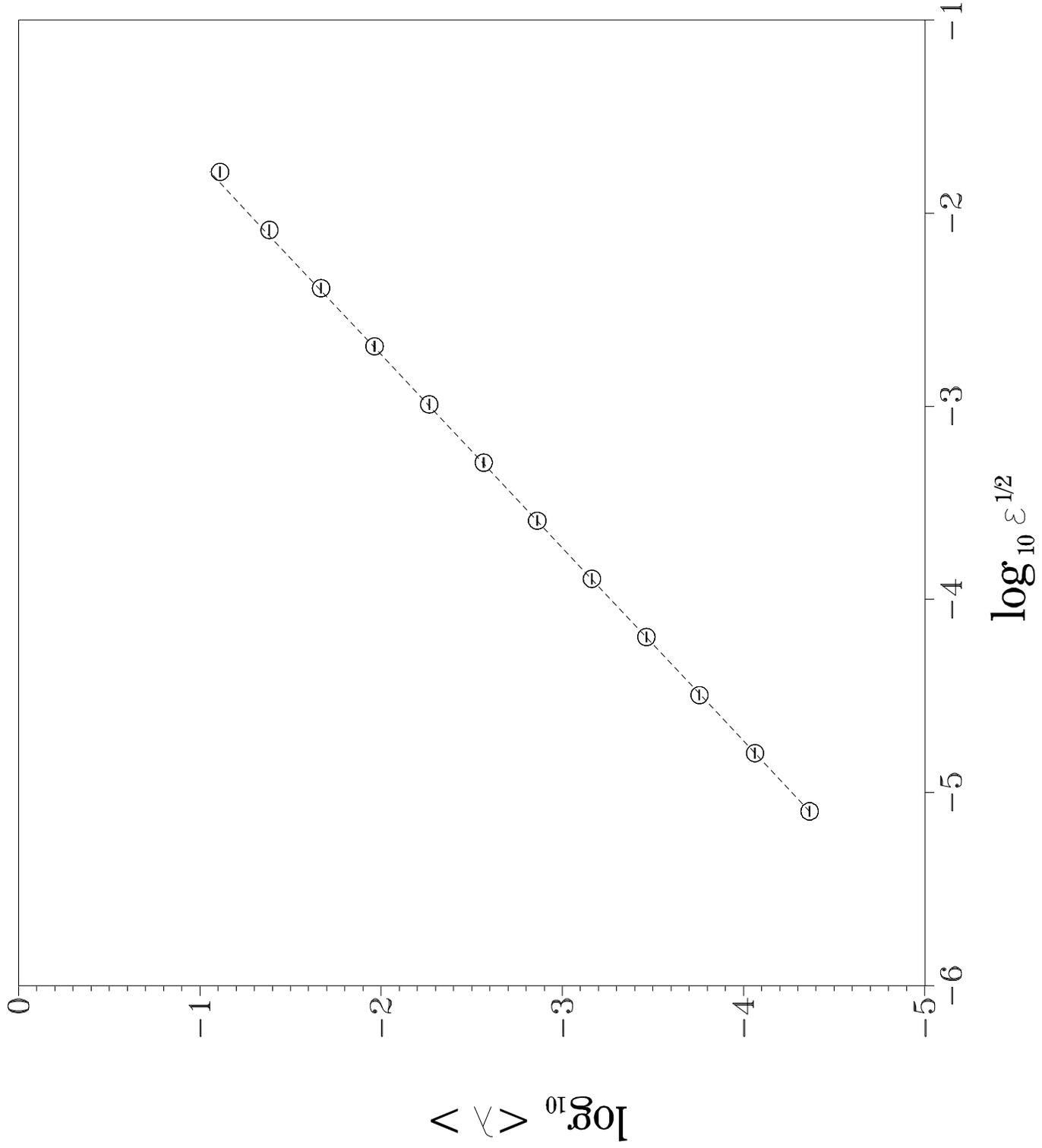
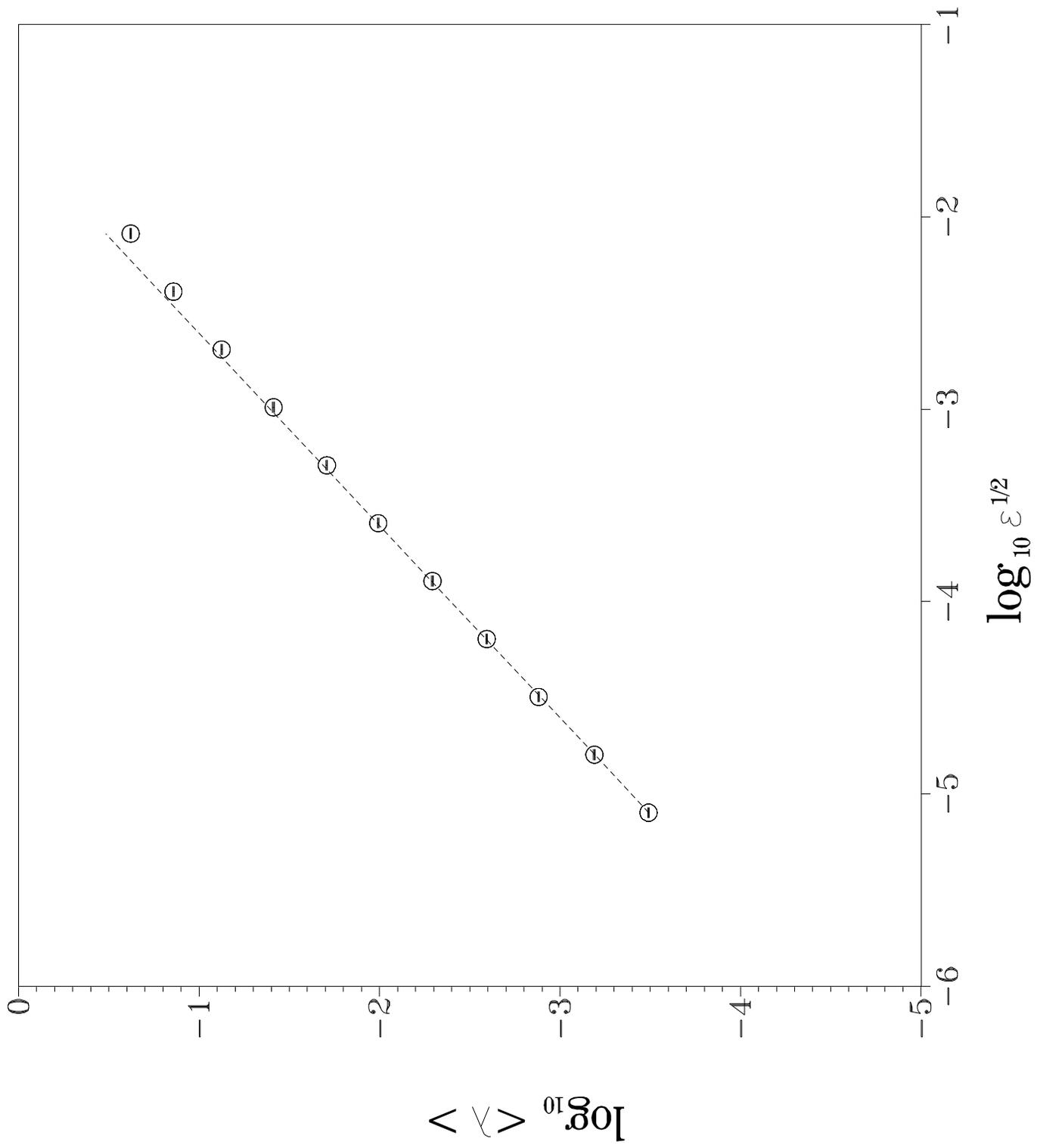
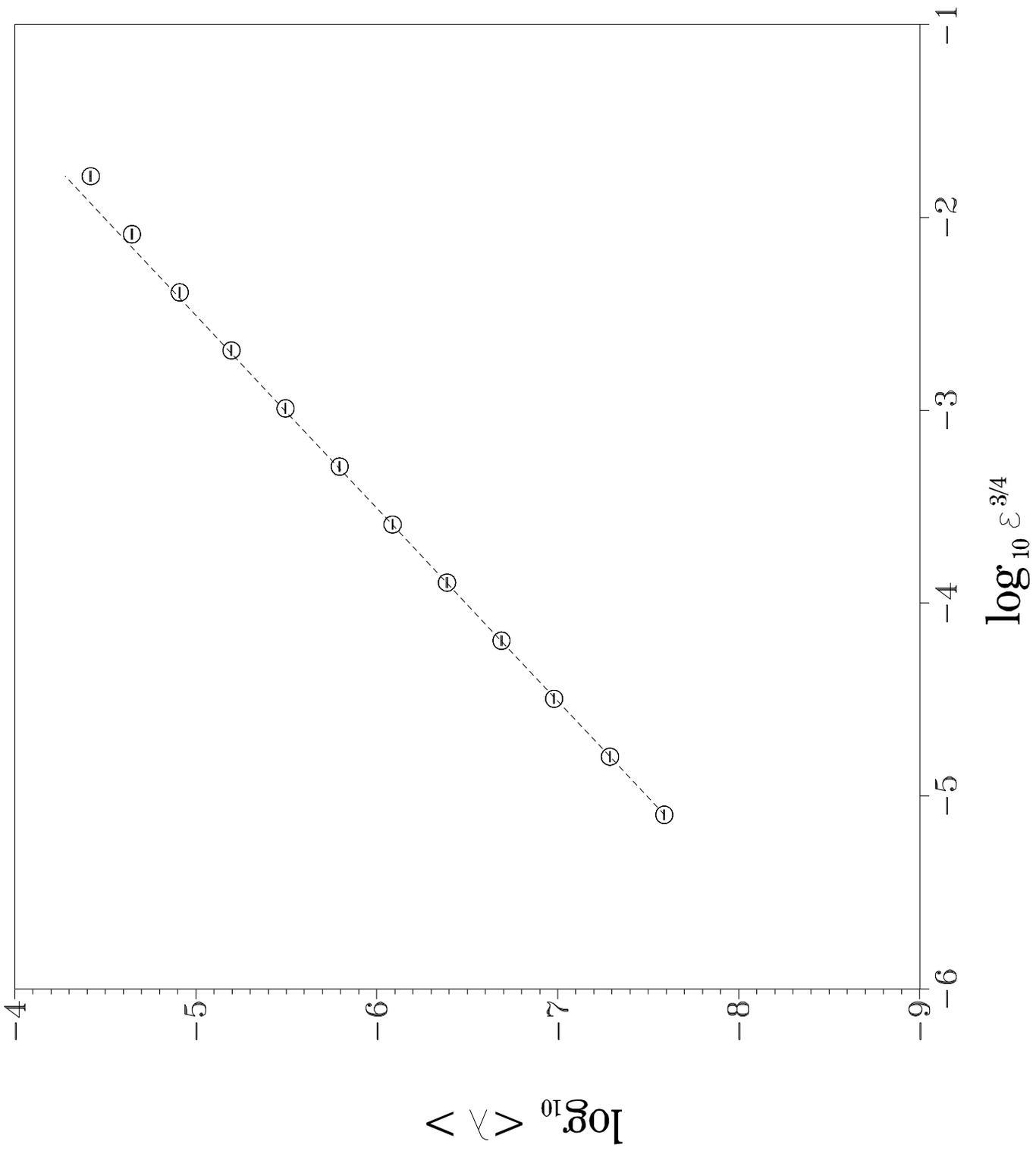
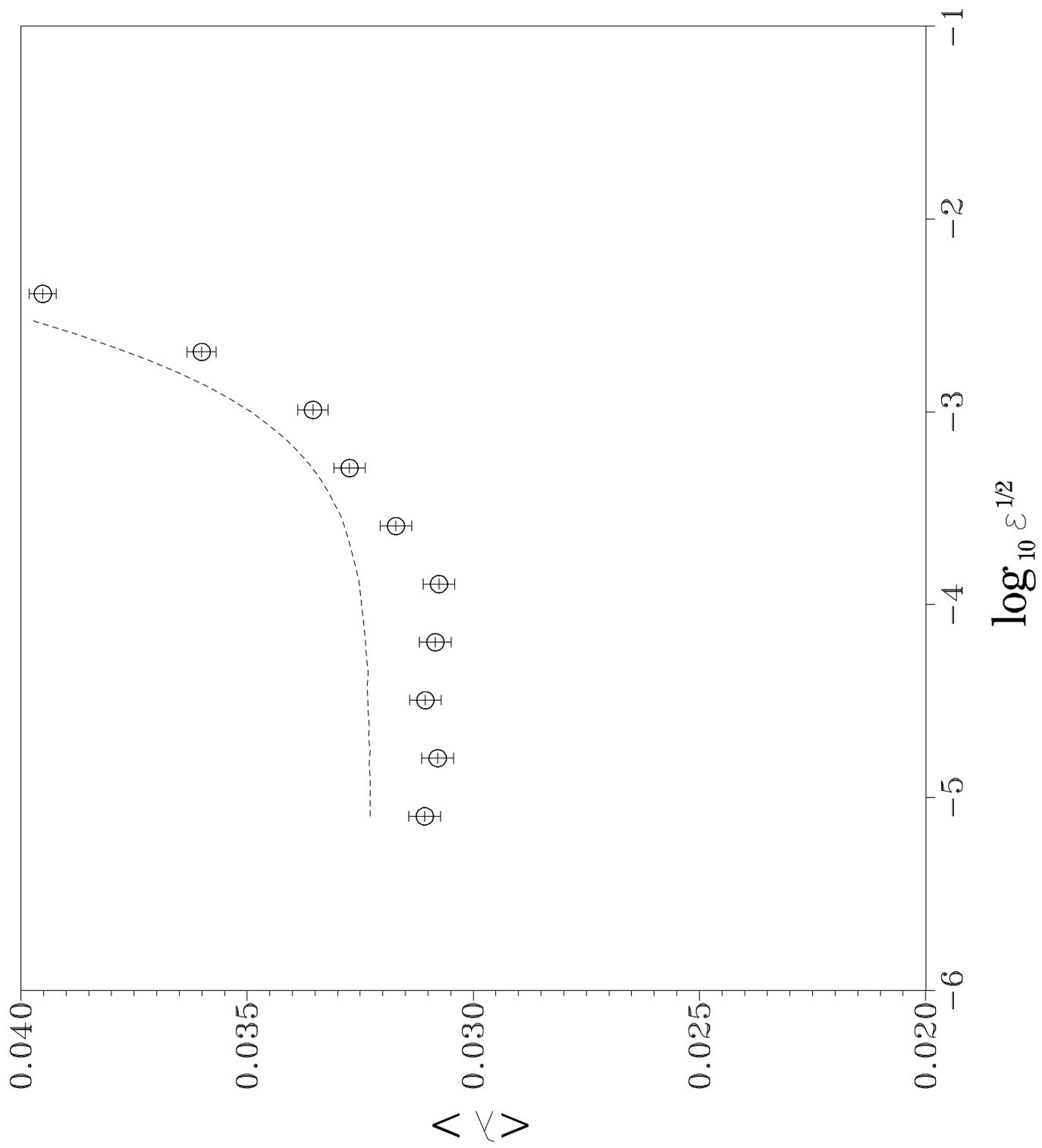
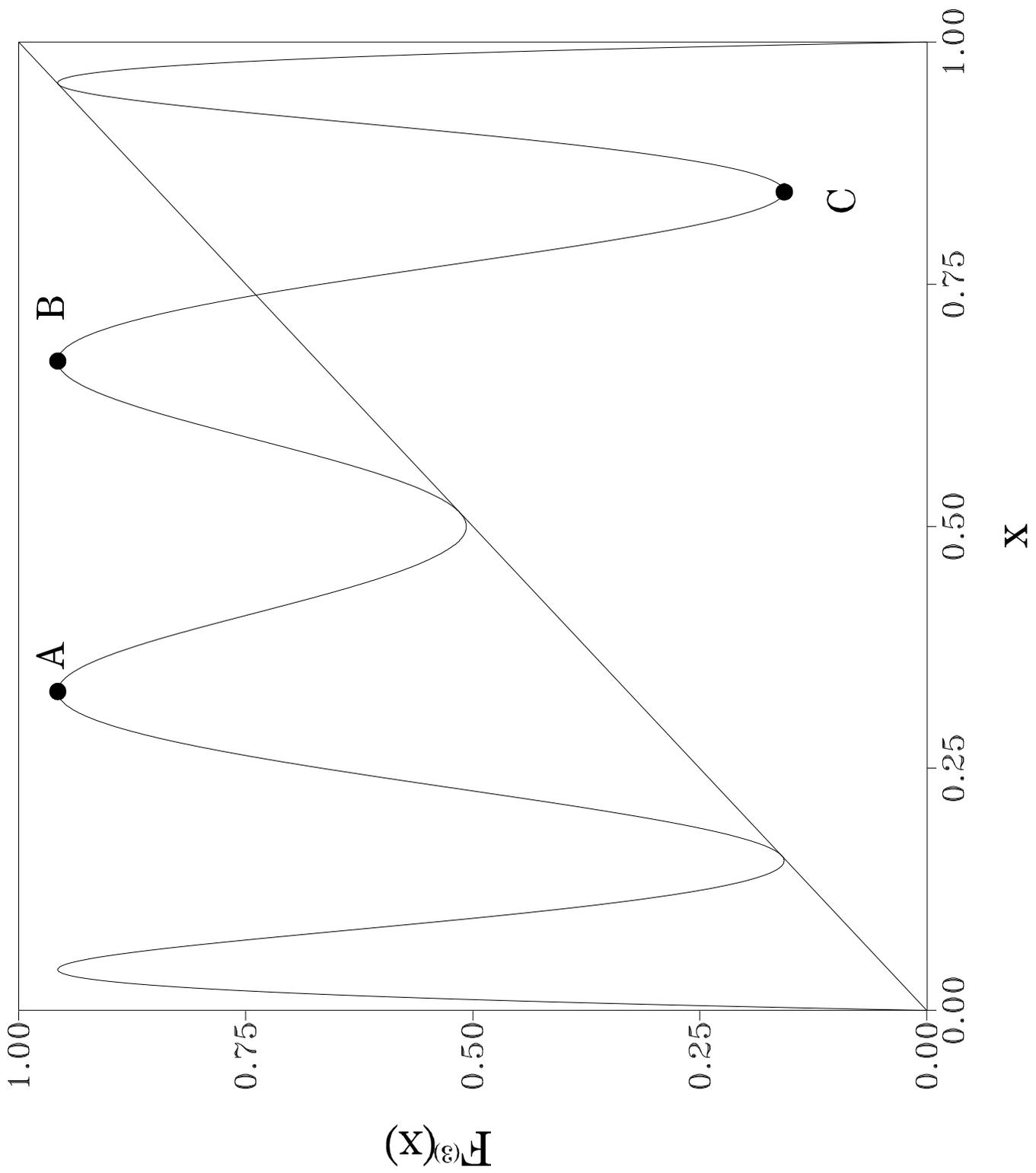
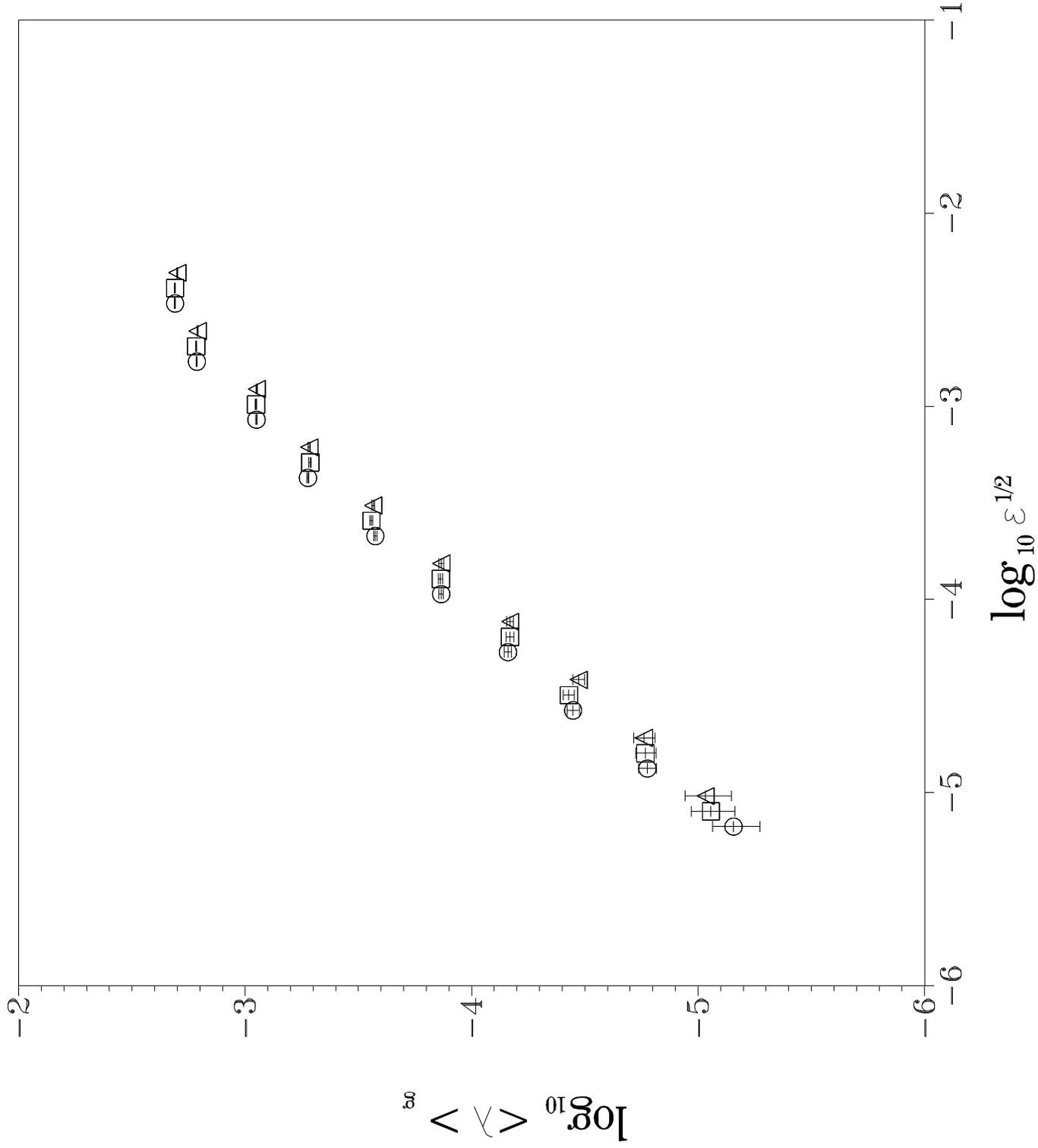
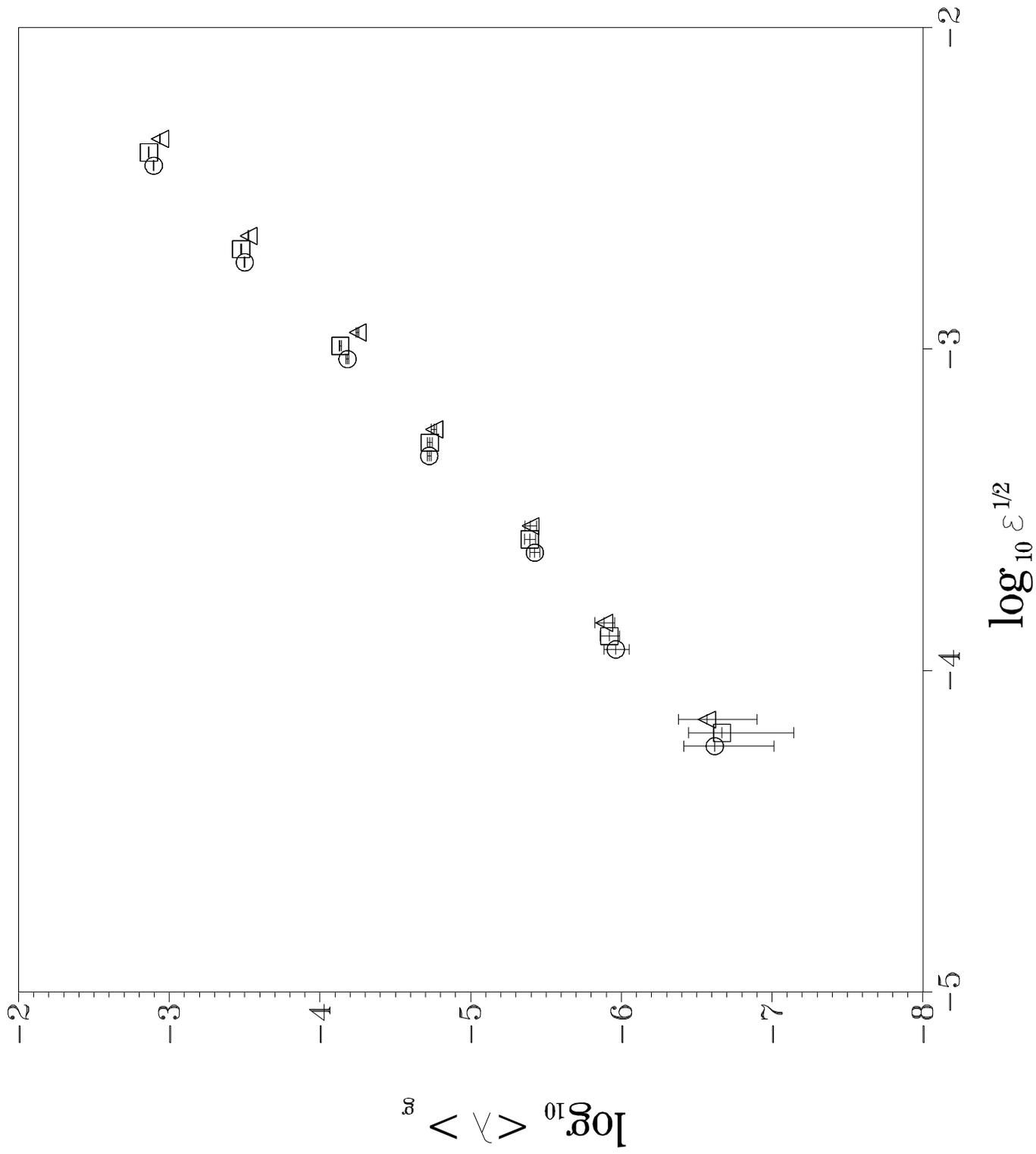
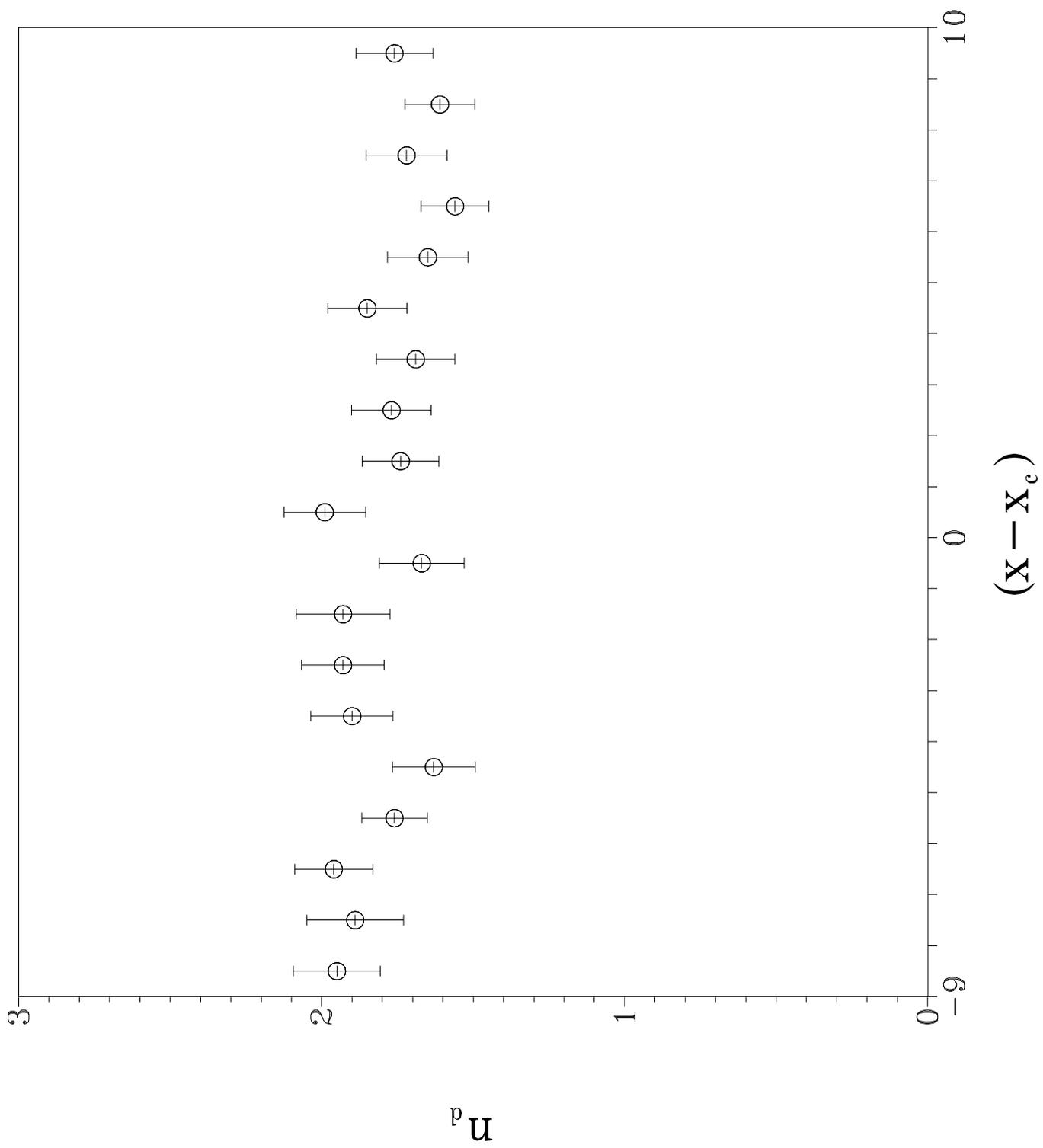
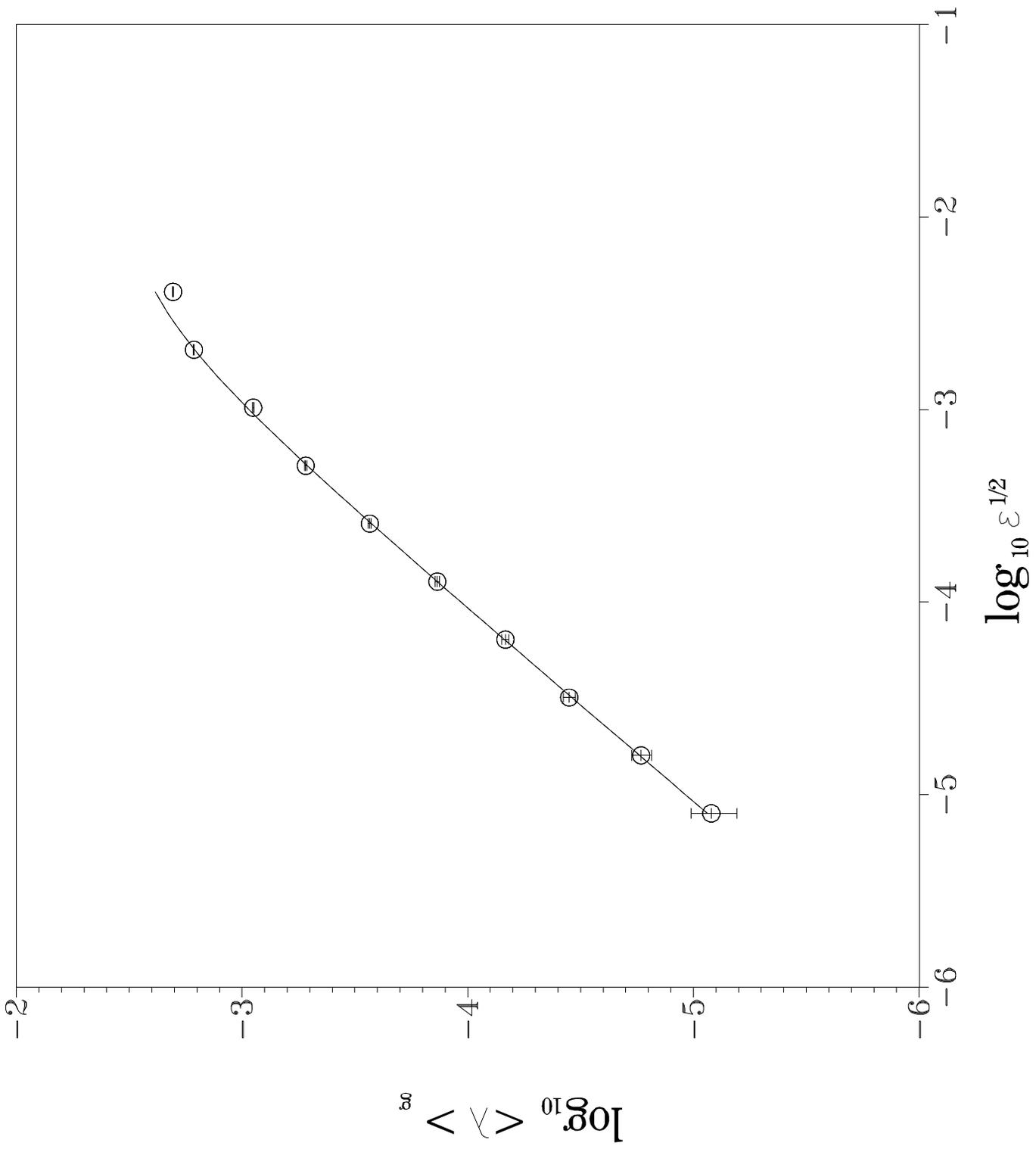
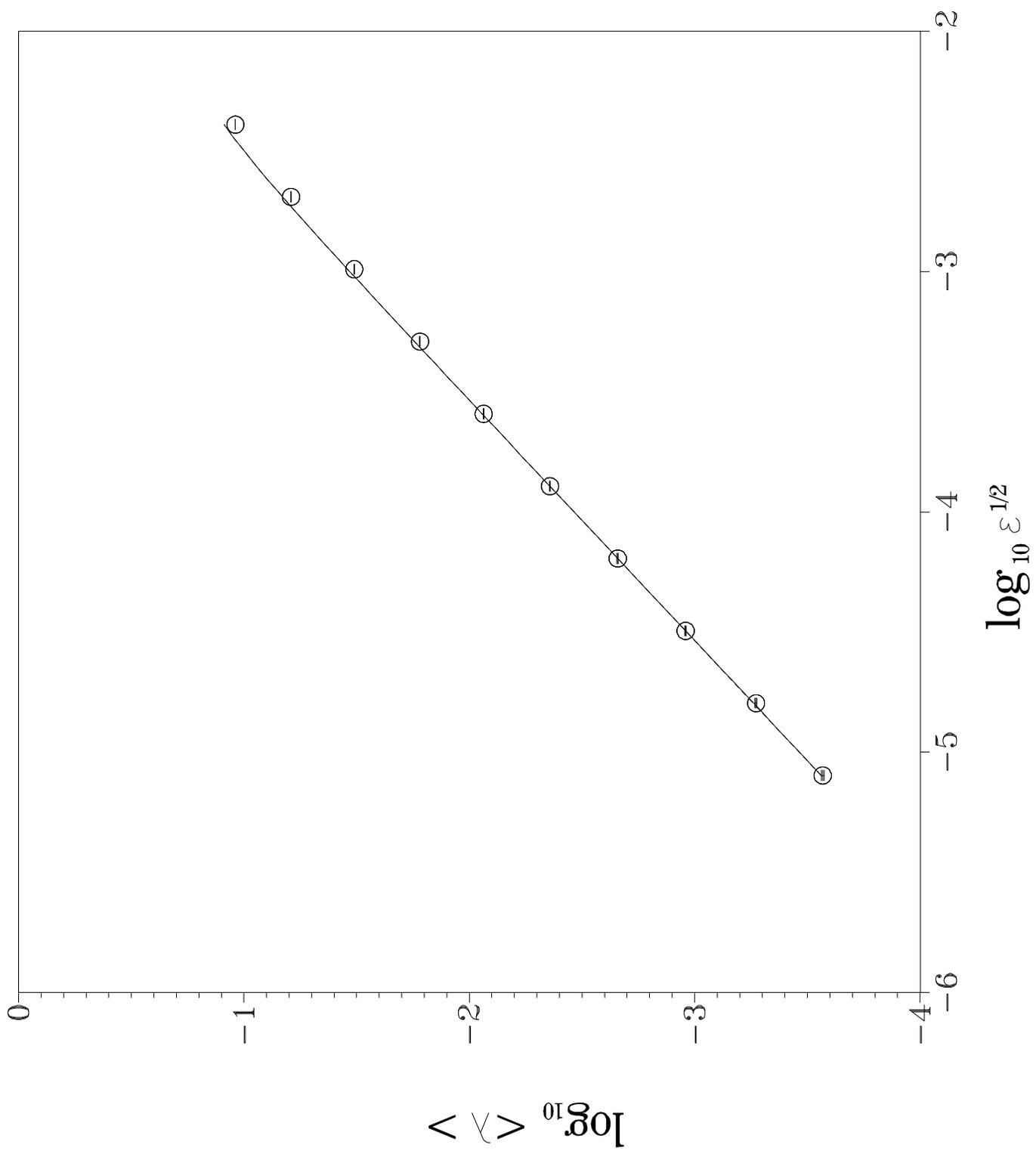
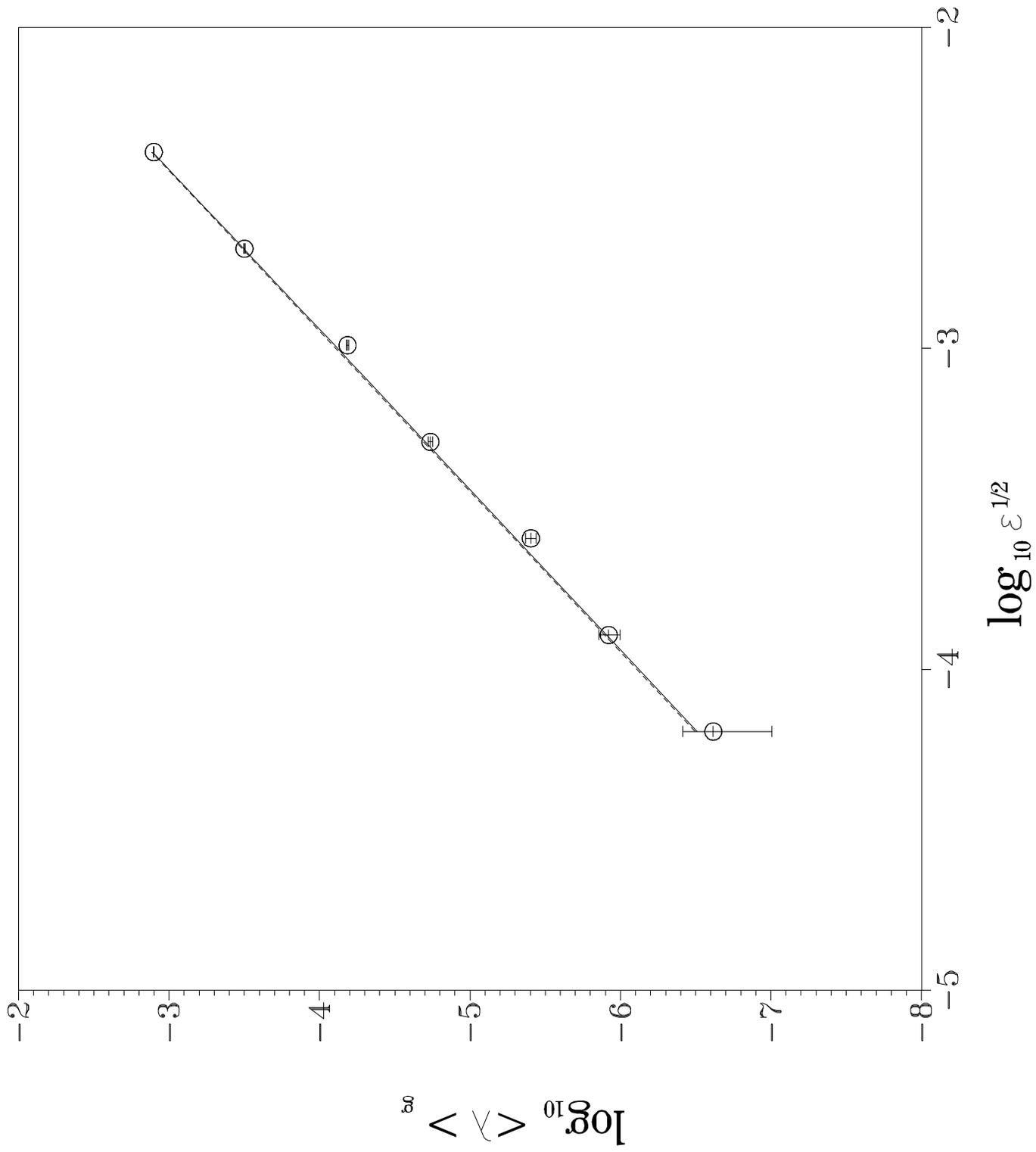
\end{document}